\documentclass[11pt,preprint]{aastex}
\newcommand{\AlIII}{\mbox{Al\,{\sc iii}}}
\newcommand{\CIV}{\mbox{C\,{\sc iv}}}

\newcommand{\FeII}{\mbox{Fe\,{\sc ii}}}
\newcommand{\feka}{\mbox{Fe\,K$\alpha$}}

\newcommand{\OVI}{\mbox{O\,{\sc vi}}}
\newcommand{\MgII}{\mbox{Mg\,{\sc ii}}}

\newcommand{\NV}{\mbox{N\,{\sc v}}}

\newcommand{\SiIV}{\mbox{Si\,{\sc iv}}}
\newcommand{\OIII}{\mbox{[O\,{\sc iii}]}}
\newcommand{\Lya}{\mbox{Ly$\alpha$}}
\newcommand{\kms}{\mbox{\,km\,s$^{-1}$}}
\newcommand{\cmsq}{\mbox{\,cm$^{-2}$}}

\newcommand{\flux}{\mbox{\,erg~cm$^{-2}$~s$^{-1}$}}
\newcommand{\flam}{\mbox{\,erg~cm$^{-2}$~s$^{-1}$~\AA$^{-1}$}}
\newcommand{\flambda}{\mbox{\,erg~cm$^{-2}$~s$^{-1}$~\AA$^{-1}$}}
\newcommand{\fE}{\mbox{\,phot~cm$^{-2}$~s$^{-1}$~keV$^{-1}$}}
\newcommand{\lumin}{\mbox{\,erg~s$^{-1}$}}
\newcommand{\persec}{\mbox{\,s$^{-1}$}}
\newcommand{\hnot}{$H_{0}$}
\newcommand{\hunits}{\mbox{\kms~Mpc$^{-1}$}}
\newcommand{\msun}{\mbox{\,${M}_{\odot}$}}
\newcommand{\nh}{\mbox{${N}_{\rm H}$}} 
\newcommand{\apm}{APM~08279+5255}

\newcommand{\pgone}{PG~1115+080}

\newcommand{\pgbal}{PG~2112+059}
\newcommand{\balq}{BAL~quasar}
\newcommand{\balqs}{BAL~quasars}
\newcommand{\balQ}{BAL~Quasar}

\newcommand{\asca}{{\emph{ASCA}}}
\newcommand{\chandra}{{\emph{Chandra}}}
\newcommand{\xmm}{\emph{XMM-Newton}}

\newcommand{\rosat}{{\emph{ROSAT}}}

\newcommand{\hst}{{\emph{HST}}}

\newcommand{\aox}{$\alpha_{\rm ox}$}

\newcommand{\aoxc}{$\alpha_{\rm ox(corr)}$}

\bibpunct{(}{)}{;}{a}{}{,}

\begin{document}
 
%\received{date month year}
%\accepted{date month year}
%\journalid{number}{date month year}
%\articleid{number}{number}
%\slugcomment{submitted to: {\it The Astrophysical Journal Letters}}

\shortauthors{Gallagher et al.}
\shorttitle{Dramatic X-ray Spectral Variability in the \balQ\ \pgbal}

%%%%%%%%%%%%%%%%%%%%%%%%%%%%%%%%%%%%%%%%%%%%%%%%%%%%%%%%%%%%%%%%%%%%%%%%%%%%%%%%%

\title{Dramatic X-ray Spectral Variability of the Broad Absorption Line Quasar \pgbal}

\author{S. C. Gallagher,\altaffilmark{1,2} 
W. N. Brandt,\altaffilmark{3}
Beverley J. Wills,\altaffilmark{4} 
J. C. Charlton,\altaffilmark{3}
G. Chartas,\altaffilmark{3}
\&
A. Laor\altaffilmark{5} 
}
 
\altaffiltext{1}{Center for Space Research, Massachusetts Institute of
Technology,  77 Massachusetts Avenue, Cambridge, Massachusetts 02139,
USA, {\em scg@space.mit.edu}}

\altaffiltext{2}{Current address: Division of Astronomy \&
  Astrophysics, University of California -- Los Angeles, 405 Hilgard
  Avenue, Los Angeles, CA 90095, USA}

\altaffiltext{3}{Department of Astronomy \& Astrophysics, The
Pennsylvania State University, University Park, PA 16802, USA}

\altaffiltext{4}{Department of Astronomy, University of Texas at Austin, Austin, 
TX 78712, USA}

\altaffiltext{5}{Physics Department, Technion, Haifa 32000, Israel}

\begin{abstract}
With a 1999 \asca\ observation, \pgbal\ became
notable as the first Broad Absorption Line (BAL) quasar
found to exhibit a typical radio-quiet quasar X-ray continuum
underlying a large amount of intrinsic absorption.  We present
a recent \chandra\ ACIS-S3 observation of \pgbal\ that demonstrates
remarkable spectral and luminosity variability since that time.
In addition to a decrease in the continuum normalization by a factor
of $\sim3.5$, the absorption column density has apparently increased
substantially, and a strong feature in the \feka\ region has appeared.
Concurrent \hst\ STIS data compared with archival \hst\ data
from earlier epochs show evidence for variability of the continuum
(up to a factor of $\sim1.7$ in the ultraviolet), and in some absorption
features of the \CIV$\lambda$1549 BAL since 1992; however, the \OVI\ BAL structure
is consistent with a 1995 observation.  We also present evidence for
Ly$\beta$--\OVI$\lambda$1037.62 
and Ly$\alpha$--\NV$\lambda$1242.80 line-locked absorption systems,
supporting the assumption that ultraviolet line pressure is driving the BAL outflow.
Whereas ultraviolet BALs
typically exhibit only modest equivalent-width variability over timescales of years, 
the dramatic X-ray variability of \pgbal\ suggests that X-ray
spectral variability studies of \balqs\ have great potential for
probing the physics of quasar winds.
\end{abstract}

%\clearpage
%%%%%%%%%%%%%%%%%%%%%%%%%%%%%%%%%%%%%%%%%%%%%%%%%%%%%%%%%%%%%%%%%%%%%%%%%%%%%%%%%

\section{Introduction}
\label{sec:intro}

Recent studies of the strong correlations between supermassive black
hole (SMBH) masses and the properties of their hosts' galactic bulges
\citep[e.g.,][]{FerMer2000,GebhardtEtal2000} demonstrate clearly the
intimate interrelation between the growth of SMBHs and their
host galaxies.  This connection implies a physical mechanism for
regulating the coeval development of these structures.
During much of their accretion phases, SMBHs
reveal themselves as luminous quasars
\citep[e.g.,][]{YuTre2002}, and thus quasar winds 
are likely to provide an important source of feedback during the coeval
growth of SMBHs and their hosts' galactic bulges \citep[e.g.,][]{Fabian1999}.
These winds are directly observed in the population of 
Broad Absorption Line (BAL) quasars, 
which exhibits deep, broad absorption lines from high-ionization
ultraviolet (UV) resonance transitions. BALs arise when our line of 
sight passes through thick material flowing outward from the
nuclear region at speeds up to 0.03--0.3$c$.  Though \balqs\ typically
compose $\sim10\%$ of the population in optically selected surveys, 
a recent analysis of the survey criteria for the Large Bright Quasar
Survey, taking into account selection biases against the \balq\ population,  
concluded that the true fraction of \balqs\ in the radio-quiet quasar population is $22\pm4\%$
\citep{HewFol2003}; \citet{ReichardEtal2003} reached a similar conclusion based on the Sloan
Digital Sky Survey \balq\ sample. Furthermore, spectropolarimetry, statistics from
gravitational lensing, and theoretical arguments support this higher
intrinsic \balq\ fraction \citep[e.g.,][]{Goodrich1997,Chartas2000,KrVo1998}.
Therefore, \balqs\ are not exotic objects but rather 
provide an opportunity to study powerful gas outflows that are
likely present in typical quasars \citep[e.g.,][]{WeMoFoHe1991}.

The presence of UV BALs is generally accompanied by extreme weakness
in soft X-rays as a result of
intrinsic absorption  with spectroscopically measured column densities
typically in the range of (1--50)$\times10^{22}$~\cmsq\ \citep[e.g.,][]{GreenEtal2001,GaBrChGa2002}.
Conversely, X-ray-weak Active Galactic Nuclei (AGNs) often show evidence for UV absorption, though not
necessarily BALs \citep*[e.g.,][]{BrLaWi2000}.
While significant X-ray and UV absorption invariably occur in
concert, the absorber properties do not always agree.  For example, the column density
implied from the X-ray observation of the \balq\ PG~0946$+$301,
\nh\,$\gtrsim10^{24}$~\cmsq\ \citep{MathurEtal2000}, is significantly larger than
the best estimates from detailed UV absorption-line studies
\citep[\nh$\sim10^{20-22}$~\cmsq;][]{AravEtal2001}.
Furthermore, among \balqs, there are no obvious
trends relating the particulars of the UV absorption-line properties
to the magnitude of the observed
weakness in X-rays \citep{GallagherEtal2003a}.  
Such a correspondence between X-ray and UV absorption properties might
be expected based on theoretical quasar disk-wind models
\citep[e.g.,][]{MuChGrVo1995}
and also on observations of Seyfert galaxies where
high-resolution X-ray spectroscopy has revealed that the ionized
absorbers seen in the X-ray regime often
share kinematic properties with the UV-absorbing outflows
\citep[e.g.,][]{KaspiEtal2002,GabelEtal2003}. However, at present it is unclear if Seyfert
ionized-absorber outflows are distinct from or belong to a continuum
of winds that extend to the energetic BAL outflows in luminous quasars
(though see Laor \& Brandt 2002 for a discussion of this issue).
\nocite{LaoBra2002}

Though a high-resolution X-ray gratings observation of a \balq\ would
offer a direct kinematic test of the connection between the velocity
structure seen in X-ray and UV absorption, the X-ray 
faintness of this population makes such an observation expensive. 
Simultaneous X-ray/UV variability studies offer an
alternative means of investigation.  
A clear demonstration of absorption variability in one wavelength regime
without corresponding changes in the other would provide strong
evidence that the two absorbers are distinct.  Conversely,
corresponding changes in both X-ray and UV absorption would
support an X-ray/UV absorber connection.  
 Understanding the relationship between X-ray and UV absorption is essential for
constraining important physical parameters, in particular the
mass-outflow rate from the central engine.  The mass-outflow rate
determines the kinetic influence of quasar activity on star formation in
the host galaxy as well as the quasar contribution to the enrichment
of the intergalactic medium with metals.

An excellent target for a variability study is the UV-bright, \balq\
\pgbal\ which has shown significant variability in both the UV
continuum and X-ray in the past \citep{GaEtal2001a}.   
This quasar is one of the most luminous low-redshift ($z=0.466$)
Bright Quasar Survey objects
with $M_{\rm V}=-26.9$ (\hnot$=70$\hunits, $\Omega_{\rm M} = 0.3$,
and $\Omega_{\Lambda} = 0.7$ are assumed throughout), and it has
a broad-band optical polarization of $0.71\%$ \citep{BeScWeSt1990}.
\hst\ spectra revealed broad, shallow \CIV\  absorption making this a \balq\ with a balnicity index
of $\approx 2980$\kms\ \citep{JannuziEtal1998,BrLaWi2000},
well above the lower limit that defines the class \citep{WeMoFoHe1991}.
There are no spectra in the
literature that cover the \MgII\ region, and so it is not known
if \pgbal\ has broad \MgII\ absorption lines (see Boroson \& Meyers 1992 for
further discussion of \MgII\ \balqs). \nocite{BoMe1992} 
However, the lack of \AlIII\ absorption (which often accompanies
\MgII) suggests that \pgbal\ does not have low-ionization BALs.
\pgbal\ was detected in soft X-rays with \rosat\ 
\citep{BrLaWi2000} which is unusual for a \balq\
\citep{KoTuEs1994,GrMa1996}. With the 1999 Oct 30 \asca\ data, 
it was the first \balq\ shown to have
an X-ray continuum typical of radio-quiet quasars underlying intrinsic
absorption.  \pgbal\ showed significant spectral and
flux variability between the 1991 Nov 19 \rosat\ and the \asca\ observations \citep{GaEtal2001a}.

With the high X-ray flux measured with \asca\ as well as its history of
X-ray variability, \pgbal\ was a natural target for a coordinated X-ray/UV
spectral study to investigate the relationship between absorption in
the two regimes.  In this paper, we present the results from joint \chandra/\hst\
observations of \pgbal\ taken on 2002 September 1.  For reference,
the dates of the multiwavelength observations presented in this paper
are presented in Figure~\ref{fig:lc}.

%--------------------------------------------------------------------------------
\section{Observations and Data Analysis}
\label{sec:obs}

\subsection{\chandra\ Analysis}
\label{sec:xanal}

On 2002 September 1, \pgbal\ was observed with the \chandra\ {\em X-ray
Observatory} \citep{chandra_ref} ACIS-S3 detector \citep{acis_ref} in
$1/8$ subarray mode to reduce the possibility of photon
pile up.  The data were processed using the standard \chandra\ X-ray Center pipeline
software from which the level~1 events file was used.  The data were
then reprocessed using \texttt{acis\_process\_events} to remove pixel
randomization and filtered on good time intervals, good \asca\ grades
(0, 2, 4, 5, 6), and good status (ignoring the bits indicating
afterglow events, which include some source X-rays).  The source
spectrum was extracted from a circular region with a $4\arcsec$
radius; the background spectrum was taken from an encircling annulus with an
inner radius of $6\arcsec$ and an outer radius of $15\arcsec$.  For
spectral analysis, the redistribution matrix file (rmf) and auxiliary
response file (arf) were generated using the
standard CIAO 2.2 software, and the arf was then corrected for the
time-dependent degradation in low-energy quantum efficiency using
\texttt{acis\_abs}.{\footnote{See http://cxc.harvard.edu/cont-soft/software/ACISABS.1.1.html.}}

Eight keV typically defines the upper limit of the ACIS energy band
as the effective area of the High Resolution Mirror Assembly drops sharply at this energy;
however, in the 7--8~keV band, only two counts were within the source
region.  Therefore \pgbal\ is not detected above 7~keV.  We ignore data
below 0.5~keV due to calibration uncertainties.
We collected 841 photons in the 0.5--7.0~keV band for a count rate of
$(1.48\pm0.05)\times10^{-2}$~ct~\persec.  
\pgbal\ was the brightest point source detected within the
$1\farcm05\times8\farcm40$ ACIS subarray field of view; the next brightest source
was $2\farcm27$ away with 60$\pm8$ full-band counts.
 The \pgbal\ radial profile appears consistent with that of a
point source, and the \chandra\ position is within $0\farcs5$ of
the NED{\footnote{NASA Extragalactic Database,
    http://nedwww.ipac.caltech.edu.}} optical position. There were no background flares, and the quasar
showed no significant variability during the 56.868~ks observation. 

Based on our best-fitting \asca\ model, we had predicted a \chandra\ ACIS-S3 count rate of
\mbox{$\approx$9.5$\times10^{-2}$~ct~\persec}, more than a factor of six larger
than observed.  This large discrepancy is much greater
than the \asca/ACIS cross-calibration uncertainties; \asca\ and ACIS-S3
fluxes agree to within $10\%$ in the 1--10~keV bandpass
\citep[e.g.,][]{Snowden2002}.  Therefore, \pgbal\ varied
significantly between the \asca\ and \chandra\ observations.
Figure~\ref{fig:chandra}a illustrates this point with the
best-fitting \asca\ model convolved with the ACIS-S3 reponse
overplotted on the current ACIS-S3 data.  Relative to the \asca\
model, the \chandra\ spectrum appears flatter 
across the \mbox{0.5--7.0~keV} band pass.  Quantitatively, 
the gross spectral shape can be approximated with a $\Gamma\sim0.75$ power law 
with the exception of a feature near the \feka\ line energy of 6.4~keV
evident in the ratio plot. 

As a first attempt to fit the \chandra\ data, we used the X-ray
spectral-fitting software XSPEC 11.2 \citep{xspec_ref} and applied the
following model from the 1999 \asca\ analysis \citep{GaEtal2001a}: 
Galactic absorption (\nh\,$=6.26\times10^{20}$~\cmsq), intrinsic neutral absorption
(\nh\,$=[0.86^{+0.54}_{-0.48}]\times10^{22}$~\cmsq), and a power law
($\Gamma=1.95^{+0.25}_{-0.24}$).\footnote{{\label{qe}The \asca\ best-fitting model
  parameters have been adjusted
  from \citet{GaEtal2001a} to account for the degradation of the
  low-energy quantum efficiency of the SIS detectors.  See
  {http://heasarc.gsfc.nasa.gov/docs/asca/calibration/nhparam.html}
  for further details.}}
Even allowing the normalization to vary freely 
did not provide a statistically acceptable fit
($\chi^{2}/\nu=168.4/47$).  The intrinsic neutral column density as
well as the power-law photon index were then allowed to vary.  Again, the 
fit was unacceptable ($\chi^{2}/\nu=84.6/47$) with clear systematic residuals near
the \feka\ region and below $\sim1$~keV in the observed frame.  
The \asca\ data could also be fit with either ionized absorber or
partial-covering neutral absorber models (though neither was
significantly preferred over a simple neutral absorber), and so we
applied the XSPEC models \texttt{absori} and \texttt{zpcfabs} 
independently, for an ionized and
partial-covering absorber, respectively.  The absorption and continuum
parameters were free to vary.  While each of these
fits offered improvements over a simple neutral absorber
($\chi^{2}/\nu=61.6/46$ and $\chi^{2}/\nu=56.5/46$ for ionized
and partial-covering absorption, respectively), clear positive
residuals near the \feka\ region indicated that another component was
required. We thus added a cosmologically redshifted gaussian emission line which
improved the $\chi^2$ value significantly for both models.  The
$\Delta\chi^2$ for three additional parameters was 12.5/12.8 for the
ionized/partial-covering absorber models.  In both
cases, this indicated a statistically significant improvement
according to the $F$-test at the $>98\%$ confidence level.  Given the
final fit statistics, we slightly prefer the partial-covering absorber model.  
However, we caution that given the data quality, the best-fitting model should be taken 
as a basic parameterization of the spectrum rather than as a truly
physical model.  In fact, the intrinsic absorber is likely to be much more
complex, and physical effects such as velocity dispersion and
multi-zone ionization structure should be kept in mind.  The \chandra\
data, best-fitting model, and residuals are presented in Figure~\ref{fig:chandra}b.

At the $90\%$ confidence level, taking all parameters to be of interest other than absolute
normalization, the best-fitting continuum model parameters are the following:
$\Gamma=2.09^{+0.75}_{-0.69}$ and 
\nh\,$=(7.03^{+3.77}_{-3.09})\times10^{22}$\cmsq, with an absorber
covering fraction of $f_{\rm cov}=0.87^{+0.10}_{-0.27}$.
The rest-frame emission-line parameters are not as well constrained, and so we
quote these errors at the $68\%$ confidence level:
$E_{\rm line}=6.38^{+0.42}_{-0.43}$~keV, 
$\sigma_{\rm line}=0.46^{+0.73}_{-0.32}$~keV, and ${\rm EW}=1050^{+520}_{-471}$~eV.  
Note that allowing the width of the emission line to vary versus forcing the line to be
unresolved improved the fit from $\chi^{2}/\nu=52.3/44$ to the final
value of $\chi^{2}/\nu=43.5/43$, which is
significant at the $>99\%$ confidence level.  Furthermore, ignoring
the observed-frame 4--5~keV region in the spectrum to ensure that the
emission-line feature was not skewing the continuum fits resulted in
best-fit values for $\Gamma$, \nh, and $f_{\rm cov}$ that were
consistent with those listed above.  
If this emission line were present in the 1999 \asca\ spectra 
with the same flux, it would have been
below the detection threshold (EW$_{\rm 1999}<383$~eV) due to the much higher
continuum normalization.  The best-fitting spectral parameters from
the partial-covering absorber model are listed in Table~\ref{tab1};
for comparison, the values from fitting the same model to the 1999
\asca\ data are also presented.  The $\chi^{2}$ contours of 
$f_{\rm cov}$ versus \nh\ for both epochs are shown in
Figure~\ref{fig:contours}; though the best-fitting intrinsic column
density has increased significantly from 1999, $f_{\rm cov}$ remains
consistent.  This result is upheld with simultaneous fitting of the \asca\ SIS
and \chandra\ data.

This intriguing result from the partial-covering absorber model
fitting that the intrinsic absorption column density had
increased significantly over a period of years required additional investigation.  
To that end, we examined whether a simple change in ionization
parameter rather than column density could account for the spectral changes
by simultaneously fitting the ACIS-S3 and \asca\ SIS0 and SIS1 data
with the \texttt{absori} model.  
We excluded the \asca\ GIS data from the fits as they are not
sensitive below $0.9$~keV, the energy regime most important for
constraining absorber properties.  Given that the photon index was consistent
between the two epochs, $\Gamma$ was tied together for all of the
subsequent model fitting. The \chandra\ data between 4--5~keV were ignored to avoid
confusion from the Fe feature.  For completeness, both the intrinsic column density and the
ionization parameter, $\xi=L_{\rm X}/(nR^{2})$~erg~cm~\persec, were
first tied together, and only the
continuum normalization was allowed to vary.  This resulted in an
unacceptable fit, with $\chi^{2}/\nu$=151.5/117.  For Model 1, 
$\xi$ was freed to vary between the \chandra\ and \asca\ epochs with a
resulting fit statistic of $\chi^{2}/\nu$=131.1/116.  The residuals
for the \asca\ data were systematically negative below $\sim1$~keV,
and the best-fitting photon index was unusually flat for a luminous quasar: $\Gamma\sim1.5$. 
For the subsequent fit (Model~2), the values for $\xi$ were tied together for both epochs, and the
column density was allowed to vary between observations.  This resulted
in an improvement to $\chi^{2}/\nu$=123.2/116 with a more plausible best-fitting
$\Gamma\sim1.7$.  Finally, for Model~3, both $\xi$ and \nh\ were allowed to vary
independently for each epoch with a resulting fit statistic of
$\chi^{2}/\nu$=112.0/115.  
The best-fitting parameters from this joint fitting for Models 1, 2,
and 3 are presented in Table~\ref{tab2}.
The substantial and significant increase in the intrinsic column
density from 1999 to 2002 implied by the independent partial-covering
absorber fitting is also supported by this ionized absorber joint modeling.   
The change in the low-energy spectrum is not consistent with a pure variation in ionization
parameter.  A simple increase in column density is acceptable
statistically, though
changes in both column density and ionization parameter are
preferred.  Interestingly, with Model~3 the \chandra\ data require a much more
highly ionized absorber with $\xi\sim500$~erg~cm~\persec\ than the
\asca\ data, which are consistent with a neutral gas.  Given that the X-ray
continuum normalization was significantly higher in 1999, this is
contrary to expectation if the X-ray continuum tracks the
ionizing continuum; the absorber would be expected to be more
highly ionized in the presence of a stronger ionizing continuum.  This
suggests that a simple ionized absorber is not appropriate in this case.

The flat spectrum with an effective $\Gamma\sim0.75$ 
and strong \feka\ emission are reminiscent of the reflection-dominated
spectra often seen from Seyfert~2 galaxies.  However, complex intrinsic absorption
can fully account for the spectral shape; the
best-fitting photon index, $\Gamma=2.09^{+0.75}_{-0.60}$, though
loosely constrained, is typical of a direct radio-quiet quasar continuum and consistent with
the 1999 \asca\ data ($\Gamma_{1999}=1.96^{+0.43}_{-0.29}$).
Furthermore, the latest self-consistent, constant-density, ionized reflector models
\citep{RosFab1993,BaIwFa2001} could not provide acceptable fits to
both the continuum and \feka\ emission feature at the same time.
Therefore, in the hard band we are either directly observing the continuum in a low
state or scattered emission from an efficient ionized ``mirror'' that preserves
the continuum shape.  

\subsection{STIS Analysis}
\label{sec:UV}
UV spectra of \pgbal\ from 1140--3170~{\AA} were obtained on 2002 September 1 with
\hst/STIS \citep{stis_ref}. Exposure times were 1100~s
and 900~s using the G140L and G230L gratings, respectively,  with
the $52\arcsec \times 0\farcs2$ slit.
The spectra have dispersions of $0.6$~{\AA}/pix and
$1.58$~{\AA}/pix, respectively, providing resolving powers of  
$R$=960--1440 for G140L and $R$=500--1010 for G230L across the full
wavelength range.  The signal-to-noise ratio  in the
continuum ranged from 10 to 25 per pixel.
Reductions were made with the standard STIS pipeline,
which provides flux calibrations accurate to $\sim3\%$ \citep{Brown2002}.
The new STIS data are overplotted on the previous \hst\ spectra in
Figure~\ref{fig:uv}.  The 1995 GHRS data covering the \OVI\ emission
and absorption features appear to be consistent with the current G140L
data, while the 1992 FOS data show a fainter and redder
continuum compared with the current G230L data.  

To investigate potential absorption variability in detail, we focused
on the \CIV, Ly$\alpha$--\NV, and \OVI\ absorption regions that are not confused
with strong emission lines as is the case for the \SiIV\ absorption region.
First, the data from the previous epochs were convolved with the
instrumental spread function and resampled to match the
lower resolution of the 2002 STIS data.  Next, wavelength shifts of $+$1,
$+$2, and $+$3 pixels were applied to the FOS G270H, FOS G190H, and
GHRS G140L spectra, respectively, to minimize the differences between previous and current
epoch spectra.  For all but the FOS G190H spectrum, these wavelength
shifts were confirmed with interstellar absorption lines.   
To investigate potential variability, the spectra from the
2002 observation were divided by those from the previous epochs (see
the middle panels of Figs.~\ref{fig:c4}--\ref{fig:o6}).  The
division plots illustrate changes in the equivalent
widths of absorption features, which should appear as significant
deviations from the mean level of the ratio plot.
The cancellation of broad emission lines in the ratio spectra in
Figures~\ref{fig:c4} and \ref{fig:o6}
indicates that the emission-line changes are consistent with their
strengths varying in proportion to the changing continuum level.

In addition, the spectra from the earlier epochs were subtracted from the
2002 spectra as shown in the lower panels of
Figures~\ref{fig:c4}--\ref{fig:o6}.  This comparison is intended to
test whether changes in absorption features seen in the ratio spectra are caused by the changing
continuum level in the presence of an unvarying component.  For
example, BAL troughs frequently contain a contribution from an
unabsorbed continuum (either from scattering or partial covering of the
primary continuum) which can fill in the troughs and cause ``non-black saturation''
\citep[e.g.,][]{Arav1997}.  In this case, an
increase in the unabsorbed continuum level while the absorbed component remained
constant could result in an apparent change in the equivalent width,
but such a change would not appear as a significant feature in the difference plot.

In Figure~\ref{fig:c4} covering the \CIV\ spectral region, the continuum level has increased
from 1992 to 2002 as can be seen clearly in the middle panel.  The ratio
of $f_{\lambda,\rm STIS}/f_{\lambda,\rm FOS}$ has a median value of
$1.50$.   In addition, positive deviations from the mean 
are evident at 1450 and 1481~\AA\ at the locations of two of the
\CIV\ absorption features (marked with asterisks in Fig.~\ref{fig:c4}).
The EWs of these two features have decreased from 1992 to 2002, at the
same time that the continuum normalization has increased.  These
concurrent changes, a decrease in absorption EW accompanied by an
increase in continuum normalization, are consistent with an increase
in ionization that could reduce the opacity of individual features.
 However, these features are not present at a comparable 
significance in the difference spectrum shown in
  the lowest panel of this figure.  This suggests that a
  constant-flux, absorbed component plus a varying, unabsorbed continuum could cause the EW changes in the troughs.
As mentioned above, this unabsorbed component at the base of the
  trough could arise when an absorber is not physically large enough to 
  cover completely the spatially extended continuum source.  The flux
  from a source other than the direct continuum, e.g., continuum flux
  scattered into the line of sight, could also fill in the absorption
  troughs.  This latter case is perhaps more likely, as an unabsorbed
  component from scattered light would be generated on
  larger spatial scales than the
  primary continuum and would therefore be expected to respond more
  slowly to changes in the primary continuum. 
  A primary continuum partially
  covered by the absorber should rise and fall in concert with
  continuum changes.

 The \CIV\ absorption feature seen at rest-frame 1450~\AA\ in
 Figure~\ref{fig:c4} has the same redshift as the 1159/1163~\AA\
\NV\ doublet marked with an asterisk in
Figure~\ref{fig:n5}.  This \NV\ component shows similar behavior: 
an apparently significant EW change is seen in the ratio spectrum without a
  correspondingly significant change in the difference plot.  Unfortunately, the \NV\
  absorber corresponding to the \CIV\ absorber at 1481~\AA\ is not
  covered by the data in Figure~\ref{fig:n5}.

Figure~\ref{fig:o6} with the same plots comparing the \OVI\ spectra from 1995 and
2002 does not demonstrate similarly significant deviations in the
ratio spectrum.  Though the continuum level has decreased by $\lesssim25\%$
blueward of the \OVI\ emission line, the absorption structure appears
consistent between the two epochs, within the uncertainties.
If the variability of the \CIV\ absorption EW does reflect actual
absorber changes, the constant level of the
\OVI\ absorption coupled with the changes in the \CIV\ structure could
indicate absorption variability between 1992 and 1995 with none
between 1995 and 2002.  
However, the difference in ionization parameter probed by
\CIV\ and \OVI\ could also allow a change in \CIV\ absorption with little
accompanying change in \OVI, depending on the ionization structure of
the absorbing gas. Unfortunately, the lack
of concurrent spectral coverage of both \CIV\ and \OVI\ in the earlier epochs
does not allow us to distinguish between these two possibilities.
Furthermore, we have no information on the properties of the UV absorber in
1999 when the \asca\ data were taken.

 In order to investigate the velocity structure in these
  absorbing features, we also examined the GHRS and FOS G190H data at higher
  resolution. 
The GHRS data have 0.80~\AA\ resolution and 0.143~\AA/pixel, and so the data were binned by
5 pixels to increase the signal-to-noise ratio, 
still giving better resolution than STIS (a 2-pixel resolution of 1.2~\AA\ for the
STIS G140L grating).  The FOS data have not been rebinned.
The wavelength scale of the GHRS data was corrected by $+$1.00~\AA, a shift
derived from the GHRS wavelength calibration observation.  This gave excellent agreement with
the expected wavelengths of several interstellar lines.

Attributing the four dips in the broad \CIV\ absorption trough to four
\CIV$\lambda\lambda1548.20,1550.77$
absorption systems leads to the identification of four
Ly$\beta\lambda$1025.72 -- \OVI$\lambda\lambda1031.93,1037.62$ systems at the same
redshifts in the GHRS spectrum.  The measured redshifts of the absorption
systems could then be refined using the higher resolution of the GHRS
spectrum, giving $z$=0.371, 0.387, 0.402, and 0.418.  These are given
in Table~{\ref{tab3}} together with estimated 1$\sigma$ uncertainties and 
outflow velocities. Figure~{\ref{fig:line-locking}}
shows details of the Ly$\beta$ -- \OVI, Ly$\alpha\lambda1215.67$ --
\NV$\lambda\lambda1238.82,1242.80$, and \CIV\
absorbed regions.  We show the positions of absorption lines
corresponding to the above four
redshifts,  plotting the data on the same logarithmic
wavelength scale so that the four systems show the same relative displacements in the
three wavelength regions.  The predicted wavelengths for Ly$\alpha$ and \NV\ are indicated in the
middle panel.

The most striking wavelength coincidences of these four systems, as
shown in the middle panel of Figure~{\ref{fig:line-locking}}, are the
overlap of \NV$\lambda$1242.80 in system 1 and  \Lya\
in system 3.  The same overlap is also seen with these lines in systems 2 and 4.
System 1 is clearly present in the Ly$\alpha$ -- \NV\ region, while
the other systems are less obvious, though all four are evident in
\OVI\ absorption in the top panel.  
In the Ly$\beta$ -- \OVI\ region, \OVI$\lambda$1031.93 of system 1
appears to coincide with Ly$\beta$ of system 2,  and
\OVI$\lambda$1031.93 of system 3 may also overlap Ly$\beta$ of system
4.

This type of coincidence, known as absorption line locking, occurs
when the relative Doppler shifts of gas at different distances from
the continuum source are
the same as the wavelength separation of two strong absorption lines.
These systems then become locked into this velocity separation, as the
system closer to the continuum source absorbs the photons that would continue
to accelerate the more distant, higher velocity system \citep*[e.g.,][]{ScaCarNoe1970,MusSolStr1972,Scargle1973}.
We suggest here that these outflowing systems
are locked in relative velocity by the coincidence of \NV$\lambda$1242.80 and Ly$\alpha$
lines, and possibly \OVI$\lambda$1037.62 with Ly$\beta$.
Ly$\alpha$--\NV\ line-locking of sub-troughs of \balqs\ has been
discussed by \citet{KoVoMoWe1993} and earlier, for example, by
\citet{TuGrFoWe1988} and \citet{WeMoFoHe1991}. If this phenomenon is
real it implies that the gas is physically stratified (e.g., system 1
is more distant than system 3), material within the BAL region is
accelerating, and that radiation pressure plays
an important role in the dynamics of the outflowing gas \citep{KoVoMoWe1993,Arav1996}.

The relative absorption strengths of the \OVI\ doublets in systems 2, 3, and possibly 1,
as well as the flat bottoms of the \CIV\ troughs which are only
partially resolved by FOS, suggest that these doublets are close to saturation.
Further evidence for saturation may come from the Ly$\beta$/Ly$\alpha$
absorption ratio in system 1, which appears much closer to unity than to
the optically thin ratio of 0.16.
Given that these absorption troughs are not black, it is probable that
there is a scattered-light component of
the continuum in the \OVI\ region, consistent with the
results from the absorption-line variability analysis above.
Higher resolution and signal-to-noise ratio data could well clarify the relationships
between the wavelengths and strengths in these UV absorption-line
systems.
Spectropolarimetry would be invaluable for differentiating between the
partial covering and scattered continuum interpretation.  Scattered
light is likely to be polarized resulting in some polarization of the
overall spectrum.  In particular, one expects a rise in the polarization level at
the bottom of the UV absorption troughs as seen in a number of \balqs\
\citep[e.g.,][]{CohenEtal1995,GooMil1995,HinWil1995}.  This approach
holds promise for \pgbal\ given its significant low level polarization
in very broad band optical polarimetry \citep{BeScWeSt1990}.

\section{Results and Discussion}

Between the 1999 \asca\ spectra and the 2002 \chandra\ spectrum presented
in this paper, the X-ray spectral shape and luminosity of \pgbal\ have varied
dramatically. From the best-fitting partial-covering absorber models, 
the change in the continuum shape at low energies
can be explained by a significant increase in the intrinsic column
density (see Figure~\ref{fig:contours}), while the underlying 
X-ray power-law photon index remained constant within
the statistical errors.  This basic conclusion is not model-dependent;
jointly fitting the \asca\ SIS and \chandra\ data with
ionized-absorber models resulted in the same conclusion (see
Table~{\ref{tab2}}). Figure~\ref{fig:4models} shows the ionized and partial-covering
absorber models which best fit the data from both epochs.
While the power-law photon indices are consistent between the two
observations, the best-fitting 2002 continuum normalization is
$\sim30\%$ of the 1999 level.  The
  absorption-corrected, rest-frame 2.0--10.0 keV luminosity,
  $L_{2.0-10.0}$, thus dropped by a factor of $\sim3.7$ while the
  absorbing column density increased (see Table~{\ref{tab1}).   
In addition, structure in the \feka\ region is now evident, 
consistent with a broad, neutral emission line with rest-frame 
EW=580--1570~eV.  A feature with this flux would not have been
detectable in the 1999 \asca\ data given the high level of the continuum.

Comparison of the UV data from 2002 with the two previous epochs, 1992
and 1995, also reveals variability, but on much more modest scales
than seen in the X-ray. The \CIV\ region observed in 1992 and 2002 shows evidence for small
equivalent width decreases in individual absorption features, 
while the \OVI\ spectra taken in 1995 and 2002 do not show significant
evidence for absorption feature changes.  
Over the course of a decade, the UV continuum luminosity varied as well; 
between 1992 and 2002 the UV continuum redward of \Lya\ 
became bluer and brighter.  Most of this increase in continuum
luminosity could have occured between 1992 and 1995 as the 1995
continuum blueward of the \Lya\ emission line is within $\sim25\%$ of
the 2002 level.

\subsection{Implications of Variability for X-ray and UV Absorbers}

Given the present data quality, translating the
significant empirical changes into physical understanding is challenging.  
Unfortunately, there are no UV observations simultaneous with the 1999
\asca\ observations, and so we cannot rule out significant changes in
the UV absorption properties between 1999 and 2002.
However, the concurrent lack of UV absorption-line or
large-scale continuum variability from the 1995 GHRS to the 2002 STIS observations
is worth noting.  The comparison of these data with the current STIS data is particularly
relevant as 1995 is closer in time to the 1999 \asca\ observation and
the continuum level in 1995 is more consistent with the 2002 level.
As seen in Figure~\ref{fig:uv}, there is no evidence for
Ly edge absorption in the 2002 STIS G140L spectrum.  This puts an upper limit on the column density
of neutral hydrogen obscuring the UV continuum of $\nh\lesssim
10^{17}$~\cmsq\ or perhaps a few times larger in the case of partial covering.
However, the best-fit values for the X-ray absorbing column
densities during both the \asca\ and \chandra\ epochs are $\sim$5--6 orders of
magnitude larger.  Either a highly ionized X-ray absorber or an X-ray
absorber that does not obstruct the UV continuum could explain this
discrepancy.  

Though UV BAL studies are notoriously unreliable for constraining absorber
column density if partial covering is not properly taken into account
\citep[e.g.,][]{Arav1997}, a column density increase by a factor of
$\sim7$ as indicated by our X-ray spectral fitting would likely have
some effect on the UV absorption lines if the X-ray and UV absorption
arise in the same gas.  The isolated, large-scale X-ray variability
thus suggests the presence of an X-ray absorber distinct from the UV absorber in \pgbal. 
However, without concurrent X-ray and UV data from more
than one epoch, this claim remains speculative.  If the bulk of the X-ray absorption is distinct,
it could arise in the ``shielding gas'' postulated in the
disk-wind models of \citet{MuChGrVo1995}.  In this picture, the broad
emission lines and the BALs are produced in the same material at
radii of a few light days with the shielding gas at smaller distances from
the SMBH. This may be typical in \balqs; the recent detection of relativistically
blueshifted X-ray absorption lines in the \balqs\ \apm\ and \pgone\
(Chartas et al. 2002; Chartas, Brandt \& Gallagher 2003),
\nocite{ChBrGaGa2002,ChBrGa2003} unmatched in velocity with UV
absorption lines, also supports this picture.

The four UV absorption systems identified in \S\ref{sec:UV}
demonstrate, if the line locking is real, that radiative forces play a
role in the dynamics of the outflow.  In particular, Lyman series, \OVI,
and \NV\ line pressure was probably important for accelerating the gas, 
otherwise the systems would not have become locked when these
wavelength coincidences occured.
However, the Ly$\alpha$--\NV$\lambda$1242.80 and Ly$\beta$--\OVI$\lambda$1037.62
line pairs are only two line-locked pairs of many other
possibilities among UV resonance lines, and so they may only contribute a
small part of the total radiation force \citep[e.g.,][]{KoVoMoWe1993}.  
If we speculate that the structure seen in BAL troughs is 
caused by systems locked into a complex network by many absorption-line pairs, then 
the UV BAL velocity structure might be expected to be extremely
stable, even in the presence of a varying continuum.  Confirmation of
a change in velocity of an X-ray BAL such as those detected in \apm\ or \pgone\ 
(if line locking plays such a significant role in UV BAL structure)
would then suggest two intriguing possibilities: (1) the X-ray absorption
is occuring in gas as it is being accelerated, before the system has
become line locked, or (2) the X-ray absorbers are being driven by Compton
pressure where line locking would not occur.  In the latter case,
the X-ray absorbers would be distinct from the UV BAL gas.

\subsection{Iron Emission}

The combination of the apparent width
($\sigma=0.46^{+0.73}_{-0.32}$~keV) and equivalent width
(EW=$1050^{+520}_{-471}$~eV) derived from modeling the Fe emission line in \pgbal\ is
notable.  Intrinsically broad lines are typically associated with emission from the
inner accretion disk; however, the EWs of such features are expected
to be $\sim150$~eV \citep[e.g.,][]{MaPePiSt1992}.  Empirically, the
mean for such lines has been found to be higher, $\sim220$~eV
\citep[e.g.,][]{GeorgeEtal2000}, though
still significantly less than we observe for \pgbal.
Therefore, if the line in \pgbal\ is intrinsically broad, this is
hard to reconcile with its large EW.  Furthermore, a time-delay scenario
in which a previously strong continuum has illuminated material whose
emission is now more apparent as the continuum level has decreased is
unlikely with the small scales ($\sim$ light hours) of the inner disk.  
The Fe emission from that region would react on short time scales to a
diminution of the direct continuum, and so a broad, high-EW line
from this region would be unlikely.

In contrast, large EW Fe lines are common features of
Seyfert~2 X-ray spectra with Compton-thick obscuration to the direct
continuum \citep[e.g.,][]{GeoFab1991,MatBraFab1996,TuGeNaMu1997}.  
In this case, reflection of direct X-rays off optically
thick material usually associated with the torus 
results in a Compton-reflection hump and narrow, fluorescent \feka\ emission with large EW. 
Therefore, an alternative picture to an intrinsically broad line 
is that the apparent breadth of the line
arises from unresolved, narrow \feka\ lines from many ionization
states.  Given the limited statistics and CCD resolution, this
scenario is plausible from the positive residuals above rest-frame 6.4~keV in
Figure~\ref{fig:chandra}, 
though the apparent lack of a strong reflection continuum (see
\S\ref{sec:xanal}) is problematic unless
the optically thick reflector is highly ionized.  Given the larger
spatial scales on which the narrow emission lines are believed to
arise, a time-delay scenario makes more sense.  If this is the case,
then the current \feka\ emission is reflecting the previously strong
level of the continuum as was seen in the  
Seyfert galaxies NGC~2992 and NGC 4051 during low continuum states
\citep[e.g.,][]{WeaverEtal1996,GuainazziEtal1998}.  
In this situation, the \feka\ line strength would be expected to
diminish over time; therefore X-ray monitoring of this source has the potential to
constrain the spatial scale of any reflecting medium in this luminous quasar. 

\subsection{\boldmath\aox}
\label{sec:aox}

%-------------------------------------------------------------This is done
X-ray spectroscopy of \balqs\ has indicated that they have typical
radio-quiet quasar X-ray continua and spectral energy distributions 
in which soft X-rays are depressed due to intrinsic absorption. 
The power in the X-ray regime relative to the UV continuum is
parameterized using the quantity
\aox=0.384~$\log(f_{\nu,\rm2keV}/f_{\nu,\rm2500\AA})$, where the flux
densities, $f_{\nu}$, are measured at the rest-frame frequencies as labelled. 
To calculate $f_{\nu,\rm2500\AA}$, the flux density at
rest-frame 2200~\AA\ (observed-frame 3225~\AA) was estimated by
extrapolating a short distance ($\sim55~\AA$) 
along the continuum from the long-wavelength end of the \hst\ STIS
spectrum.  From that point, the
flux density at rest-frame 2500~\AA\ was calculated using the Sloan Digital Sky
Survey composite quasar spectral slope, $\alpha_{\nu}=-0.44$
\citep{VandenBerk2001}. We chose this extrapolation to avoid
potentially significant contamination from a complex of blended \FeII\
emission lines around 2500~\AA, and this value for the 
spectral index is consistent with the observed UV continuum.
The rest-frame 2~keV (observed-frame
1.36~keV) flux density is determined from a local continuum fit to the \chandra\ data.
Based on these simultaneous X-ray and UV data of \pgbal, the observed value of
\aox=$-2.07\pm0.05$ is within the normal range for \balqs\ \citep[$-$2.6
to $-$1.9; e.g.,][]{GallagherEtal2003a,GreenEtal2001}. However, correcting the X-ray spectrum
for the intrinsic absorption from the best-fitting partial-covering
absorber gives a ``corrected'' \aoxc=$-1.79$.
In contrast, the value calculated from the absorption-corrected 1999
\asca\ spectra and the 1992 \hst\ FOS data, \aoxc$_{1999}=-1.55$,
suggested much stronger (a factor of $\sim 4.2$) X-ray emission
relative to the UV continuum, though the 1992 UV data were not
simultaneous with the \asca\ observation.  

As can be seen in Figure~\ref{fig:uv}, the continuum flux densities at rest-frame 2200~\AA\
(observed-frame 3225~\AA) in both epochs are very similar, and so the
decrease in \aox\ is dominated by the decrease in X-ray flux.
For comparison, normal radio-quiet
quasars of this luminosity typically have \aox$\sim-1.62\pm0.1$
\citep[see  eq. 4 of][]{VignaliEtal2003b}.  The change in \aoxc\ of $-0.24$ over the
$\sim3$-yr time span is thus still within the spread for normal
radio-quiet quasars, though \aoxc$_{2002}$ is at the weak end of the distribution.
At this point, it is unclear whether the 1999 \asca\ spectra or the
2002 \chandra\ data are more representative of the ``normal'' state of the quasar
continuum in \pgbal.  If the 1999 \asca\ data are more representative, then \pgbal\ might return
to this emission level at any time.  At least one \balq, PG~1254$+$047, was argued
to be intrinsically X-ray-weak in addition to suffering from heavy
X-ray absorption \citep{SaHa2001}.  Though the X-ray data quality in that
particular case was poor (only $\sim44$ photons), it is quite possible
that the strength of the X-ray continuum of that source varies as well. 
Probing the timescale and interplay of X-ray continuum and absorption
variability requires additional monitoring observations of
sources with demonstrated variability such as \pgbal.

\subsection{Comparison of the Bolometric and Eddington Luminosities}

In contrast to the conclusion drawn by \citet{WeMoFoHe1991} that all
quasars are likely to have BAL regions, others have suggested that the
BAL phenomenon may instead be a manifestation of an evolutionary
phase, perhaps of enhanced accretion rate
\citep[e.g.,][]{HaMoTeMc1984,BeckerEtal2000}.  To investigate this
issue for \pgbal, we estimated the bolometric luminosity as a fraction
of the Eddington luminosity, $L_{\rm bol}/L_{\rm Edd}$, to determine if \pgbal\ is an outlier
in the $L_{\rm bol}/L_{\rm Edd}$ distribution of unabsorbed radio-quiet quasars. 
From $I$-band imaging, \citet{FyMoTh2001} measure $I=18.8$ for the apparent
magnitude of the luminous elliptical host galaxy which corresponds to an absolute magnitude of
$M_I=-23.2$  (in the observed-frame Cousins $I$ band).  After
converting to the rest-frame $V$-band luminosity \citep[including the appropriate
  $K$ and color-corrections for elliptical galaxies from Table 3
  of][]{FukShiIch1995}, we obtain an estimate of the black hole mass
of $M_{\rm BH}$=8.8$\times10^{8}\msun$ \citep[using eq. 7 from][]{Laor2001}.
To estimate the black hole mass from the optical spectroscopic properties
of the quasar, we utilize $L_{\rm 3000 \AA}$ and H$\beta$ FWHM
and eq. 3 in \citet{Laor1998} to
obtain $M_{\rm BH}$=6.8$\times10^{8}\msun$.  Both of these mass estimates have significant
 systematic uncertainties, perhaps as large as a factor of $\sim3$
\citep[e.g.,][]{Laor1998,Vestergaard2002}.
However, it is notable that these two completely independent methods
give comparable estimates of the black hole mass. 

 Using the bolometric correction from the
 2500~\AA\ monochromatic luminosity for
radio-quiet quasars of 6.3$\pm3.1$ \citep[as calculated empirically by][]{ElvisEtal1994}, we
estimate a bolometric luminosity for \pgbal\ of
$L_{\rm bol}=(7.7\pm3.8)\times10^{46}$~\lumin\ based on
 $L_{\rm2500\AA}$ measured from the 2002 STIS spectrum as described in
 $\S$\ref{sec:aox}.  Combining
 these values with the log average of the black hole
mass estimates, $M_{\rm BH}$=7.7$\times10^{8}\msun$, gives an
Eddington ratio of $L_{\rm bol}/L_{\rm
  Edd}\sim0.8$. Comparison of this rough value with those of
similarly luminous quasars in the sample accumulated by \citet[][see their
  Figure~8]{WooUrr2002} indicates that \pgbal\ does not have an anomalously high Eddington
 ratio at present.  Similarly, \citet{YuaWil2003} found that the
 Eddington ratios of a sample of $z\sim2$ BAL and non-\balqs\ were consistent.

%-------------------------------------------------------------This is
%done

\subsection{Future Work}

The lack of variability in the velocity structure of UV BALs on
timescales of years is one of the most serious challenges to the current disk-wind models
\citep[e.g.,][]{deKool1997}.  While changes in the depths of
individual absorption-line features have been seen, 
this EW variability is typically ascribed to changing ionization
structure within the absorbing gas \citep[e.g.,][]{BarlowEtal1992}.
Such a change in ionization is consistent with the decrease in the depth of \CIV\
absorption features we detected during an epoch when the UV
continuum increased in normalization as well as became bluer.  
In contrast, changes in velocity of individual absorption components are almost unknown.
However, the most comprehensive absorption-line variability investigation to
date, the 3-yr study by \citet{Barlow1993}, was not long enough for a
conclusive test; the hydrodynamic wind models of \citet{PrStKa2000} generated
instabilities on rest-frame timescales of $\sim3$~yr.
With longer time baselines, velocity increases have been claimed for at least
one \balq, Q1303$+$308 \citep{VilIrw2001}, and an extended study is 
warranted to investigate this issue further.  
With high-quality UV coverage now extending from 1992 to
2002, almost seven years in the quasar rest-frame, \pgbal\ is an excellent
candidate for continued monitoring of its BAL structure.

The large gaps in coverage of \pgbal\ over
the last decade, illustrated in Figure~\ref{fig:lc}, have limited our
ability to interpret its multiwavelength variability. 
Additional epochs of simultaneous X-ray/UV monitoring will contribute to
understanding the connection between X-ray and UV absorption in \balqs.
The detection of four narrow, probably line-locked absorption systems indicates
that high resolution UV spectroscopy covering these features
would enable a more detailed investigation into the partial covering
properties of these UV absorption systems, for comparison with the
partial covering of the X-ray gas.
Furthermore, if the bulk of the X-ray absorbing
gas lies at smaller radii than the UV BAL gas,
then the X-ray absorbers might be expected to vary, in ionization state,
column density, and velocity, on more rapid timescales than the UV BALs.  
There are now examples of \balqs\ with
evidence for X-ray absorption variability: \pgbal, \apm, and perhaps
\pgone\ \citep{Chartas2000,ChBrGa2003}.  
These three objects compose a significant
fraction of the handful of \balqs\ that have X-ray data adequate for
variability studies; this demonstrates that the X-ray regime may be a
fertile area for exploring details of quasar winds inaccessible in the
UV. 

Furthermore, to interpret X-ray variability accurately, the
hydrodynamic disk-wind models \citep[e.g.,][]{PrStKa2000} may need to
be extended to smaller radii ($<10^{15}$~cm) with higher resolution, 
and they must also more explicitly include the effects of
Compton and X-ray edge and line pressure.  For the highly ionized gas expected at small
radii,  these will dominate the radiation force over UV line pressure \citep{CheNet2003a,CheNet2003b}.
More appropriate models for low-energy absorbers which include the
potentially substantial opacity from velocity-broadened X-ray BALs are
also required.  This line opacity could be comparable to the
edge opacity in certain cases.
Additional items for an X-ray observer's wish list from theoretical
modeling are specific predictions for the strength, profiles, and
variability properties of iron emission and absorption lines.

\acknowledgements
We thank Andrea Garcia-Rissman for providing us with optical monitoring
data for \pgbal.  We also thank Herman Marshall and Norbert Schulz for sharing
their expertise on the cross-calibration of \asca\ and ACIS. 
Paul Green contributed valuable referee comments which improved
this paper. This work was supported by \chandra\ X-ray Center
grant \mbox{GO2--3129} (SCG, WNB), \hst\ grant
\mbox{GO--09277.01--A} (SCG, JCC), NASA LTSA grant NAG5--13035 (WNB),
NASA LTSA grant NAG5--10817 (JCC, WNB), and NSF grant \mbox{AST--0206261} (BJW).

%%%%%%%%%%%%%%%%%%%%%%%%%%%%%%%%%%%%%%%%%%%%%%%%%%%%%%%%%%%%%%%%%%%%%%%%%%%%%%%%

%%%%%%%%%%%%%%%%%%%%%%%%%%%%%%%%%%%%%%%%%%%%%%%%%%%%%%%%%%%%%%%%%%%%%%%%%%%%%%%%%
%\bibliographystyle{apj3}
%\bibliography{pg2112,refs}
%%%%%%%%%%%%%%%%%%%%%%%%%%%%%%%%%%%%%%%%%%%%%%%%%%%%%%%%%%%%%%%%%%%%%%%%%%%%%%%%%
\begin{deluxetable}{lcc}
\tablewidth{0pt}
\tablecaption{Parameters of the Partial-Covering Absorber X-ray Spectral Models
\label{tab1}}
\tablehead{
\colhead{} &
\colhead{\asca\tablenotemark{b}}         &
\colhead{\chandra}      \\
\colhead{Property\tablenotemark{a}}      &
\colhead{1999 Nov} &
\colhead{2002 Sep} 
}
\startdata
\nh\ ($10^{22}$~\cmsq) &
$0.94^{+2.16}_{-0.34}$&
$7.03^{+3.77}_{-3.09}$
\\
$f_{\rm cov}$ &
$0.91^{+0.09}_{-0.43}$&
$0.87^{+0.10}_{-0.27}$
\\
$\Gamma$ &
$1.96^{+0.43}_{-0.29}$&
$2.09^{+0.75}_{-0.69}$
\\
$N_{\rm 1~keV(obs)}$ ($10^{-5}$~\fE) &
%p. 82 in research notebook
$29.0^{+3.2}_{-3.2}$& 
%p. 78 in research notebook
$8.2^{+1.3}_{-1.0}$
\\
$\chi^2/\nu$ &
$139.8/148$&
$43.5/43$
\\
Line Energy (keV) &
$\cdots$&
$6.38^{+0.42}_{-0.43}$
\\
$\sigma$ (keV)  &
$\cdots$&
$0.46^{+0.73}_{-0.32}$
\\
EW (eV) &
$<383$\tablenotemark{c}&
$1050^{+520}_{-471}$
\\
$F_{0.5-2.0}$ ($10^{-13}$~\flux)&
$2.64^{+0.29}_{-0.30}$ &
$0.26^{+0.43}_{-0.31}$ 
\\
$F_{2.0-7.0}$ ($10^{-13}$~\flux)&
$5.84^{+0.64}_{-0.65}$&
$1.28^{+0.18}_{-0.13}$
\\
$L_{2.0-10.0}$ ($10^{44}$~\lumin)\tablenotemark{d}&
%p. 79 of research notebook
$6.08^{+0.64}_{-0.69}$&
$1.63^{+0.25}_{-0.17}$
\\

\enddata
\tablenotetext{a}{All values listed are for the rest frame, except for
  fluxes and power-law normalizations.  Errors quoted are for $90\%$ confidence taking all
  model parameters to be of interest other than absolute normalization with
  the exception of the line parameters, which are quoted for
  $68\%$ confidence.  All fits include fixed Galactic absorption,
  \mbox{\nh\,=6.26$\times10^{20}$~\cmsq}.}
\tablenotetext{b}{\asca\ best-fitting parameters have been adjusted
  from \citet{GaEtal2001a} to account for the degradation of the
  low-energy quantum efficiency of the SIS detectors (see
  $\S$\ref{sec:xanal}).}
\tablenotetext{c}{The EW upper limit from the \asca\ data 
  is calculated assuming the energy and $\sigma$ measured for the
  \chandra\ emission line.  For an unresolved neutral \feka\ emission line, the
  EW$<210$~eV \citep{GaEtal2001a}.}
\tablenotetext{d}{Rest-frame 2.0--10.0~keV luminosity corrected for
  both Galactic and intrinsic absorption.}
\end{deluxetable}
%----------------------------------------------------------------------

%\stepcounter{footnote}
%\footnotetext{See
%  {http://heasarc.gsfc.nasa.gov/docs/asca/calibration/nhparam.html}
%  for more details.}

%%%%%%%%%%%%%%%%%%%%%%%%%%%%%%%%%%%%%%%%%%%%%%%%%%%%%%%%%%%%%%%%%%%%%%%%%%%%%%%%%
\begin{deluxetable}{lcc}
\tablewidth{0pt}
\tablecaption{Ionized-Absorber Model Parameters from Joint Fitting
\label{tab2}}
\tablehead{
\colhead{} &
\colhead{\asca\ SIS}         &
\colhead{\chandra}      \\
\colhead{Property\tablenotemark{a}}      &
\colhead{1999 Nov} &
\colhead{2002 Sep} 
}
\startdata
\multicolumn{3}{c}{Model 1} \\
\nh\ ($10^{22}$~\cmsq) &
\multicolumn{2}{c}{$9.5^{+7.2}_{-5.1}$}
\\
$\xi$ (erg~cm~\persec)&
$4999^{\rm b}_{-3414}$&
$582^{+533}_{-302}$
\\
$\Gamma$ &
\multicolumn{2}{c}{$1.50^{+0.15}_{-0.16}$}
\\
$\chi^2/\nu$ &
\multicolumn{2}{c}{131.1/116}
\\
\tableline
\multicolumn{3}{c}{Model 2} \\
\nh\ ($10^{22}$~\cmsq) &
$2.8^{+2.3}_{-0.0}$&
$10.1^{+8.5}_{-5.1}$
\\
$\xi$ (erg~cm~\persec)&
\multicolumn{2}{c}{$448^{+821}_{-224}$}
\\
$\Gamma$ &
\multicolumn{2}{c}{$1.67^{+0.35}_{-0.33}$}
\\
$\chi^2/\nu$ &
\multicolumn{2}{c}{123.2/116}
\\
\tableline
\multicolumn{3}{c}{Model 3} \\
\nh\ ($10^{22}$~\cmsq) &
$0.6^{+2.6}_{-0.4}$ &
$11.9^{+7.2}_{-4.3}$
\\
$\xi$ (erg~cm~\persec)&
$0^{+120}$ &
$502^{+411}_{-214}$
\\
$\Gamma$ &
\multicolumn{2}{c}{$1.72^{+0.25}_{-0.24}$}
\\
$\chi^2/\nu$ &
\multicolumn{2}{c}{112.0/115}
\\

\enddata
\tablenotetext{a}{All absorber values listed are for the rest frame.  
Errors quoted are for $90\%$ confidence taking 3 parameters (\nh,
$\xi$, and $\Gamma$) to be of interest.  All fits include fixed Galactic absorption
  (\mbox{\nh\,=6.26$\times10^{20}$~\cmsq}), and the \asca\ SIS spectral
  models also include the corrections for the degradation of the low-energy
  quantum efficiency (see $\S$\ref{sec:xanal}).}
\tablenotetext{b}{This value for $\xi$ is ``pegged'' at the upper
  limit of the allowed range.}
\end{deluxetable}
%----------------------------------------------------------------------

%%%%%%%%%%%%%%%%%%%%%%%%%%%%%%%%%%%%%%%%%%%%%%%%%%%%%%%%%%%%%%%%%%%%%%%%%%%%%%%%%
\begin{deluxetable}{cccr}
\tablewidth{0pt}
\tablecaption{Properties of Line-Locked Absorption Systems
\label{tab3}}
\tablehead{
\colhead{System} &
\colhead{} &
\colhead{} &
\colhead{$v_{\rm outflow}$} \\
\colhead{Number} &
\colhead{$z_{\rm abs}$} &
\colhead{$\beta_{\rm outflow}$\tablenotemark{a}} &
\colhead{(\kms)} 
}
\startdata
%\multicolumn{3}{c}{Model 1} \\
%\nh\ ($10^{22}$~\cmsq) &
%\multicolumn{2}{c}{$9.5^{+7.2}_{-5.1}$}
%\\
1  &
$0.37087\pm0.00017$ &
$0.067\pm0.004$ &
$20,100\pm1300$
\\
2 &
$0.38725\pm0.00023$ &
$0.055\pm0.004$ &
$16,500\pm1300$
\\
3 &
$0.40156\pm0.00010$ &
$0.045\pm0.004$ &
$13,500\pm1300$
\\
4 &
$0.41837\pm0.00012$ &
$0.033\pm0.004$ &
$9,900\pm1300$ 
\\
\enddata
\tablenotetext{a}{The values of $\beta_{\rm outflow}=v_{\rm
    outflow}/c=[(1+z_{\rm em})^{2} -
    (1+z_{\rm abs})^{2}]/[(1+z_{\rm em})^{2} + (1+z_{\rm abs})^{2}]$ 
  were calculated using a
  systemic redshift of $z_{\rm em}$=0.466 measured by
  \citet{BorGre1992} from fitting the narrow \OIII\ emission
  line.  Note, however, that the weakness of the \OIII\ emission
 in \pgbal\ introduces some uncertainty into this redshift
 determination, and \citet{CorBor1996} measured $z_{\rm em}=0.460$
 from broad H$\beta$. The stated errors take into account the
 uncertainties in both $z_{\rm abs}$ and $z_{\rm em}$.}
\end{deluxetable}
%----------------------------------------------------------------------

%%%%%%%%%%%%%%%%%%%%%%%%%%%%%%%%%%%%%%%%%%%%%%%%%%%%%%%%%%%%%%%%%%%%%%%%%%%%%%%%
\begin{figure}[t]
\centerline{\includegraphics[scale=1.0,angle=0]{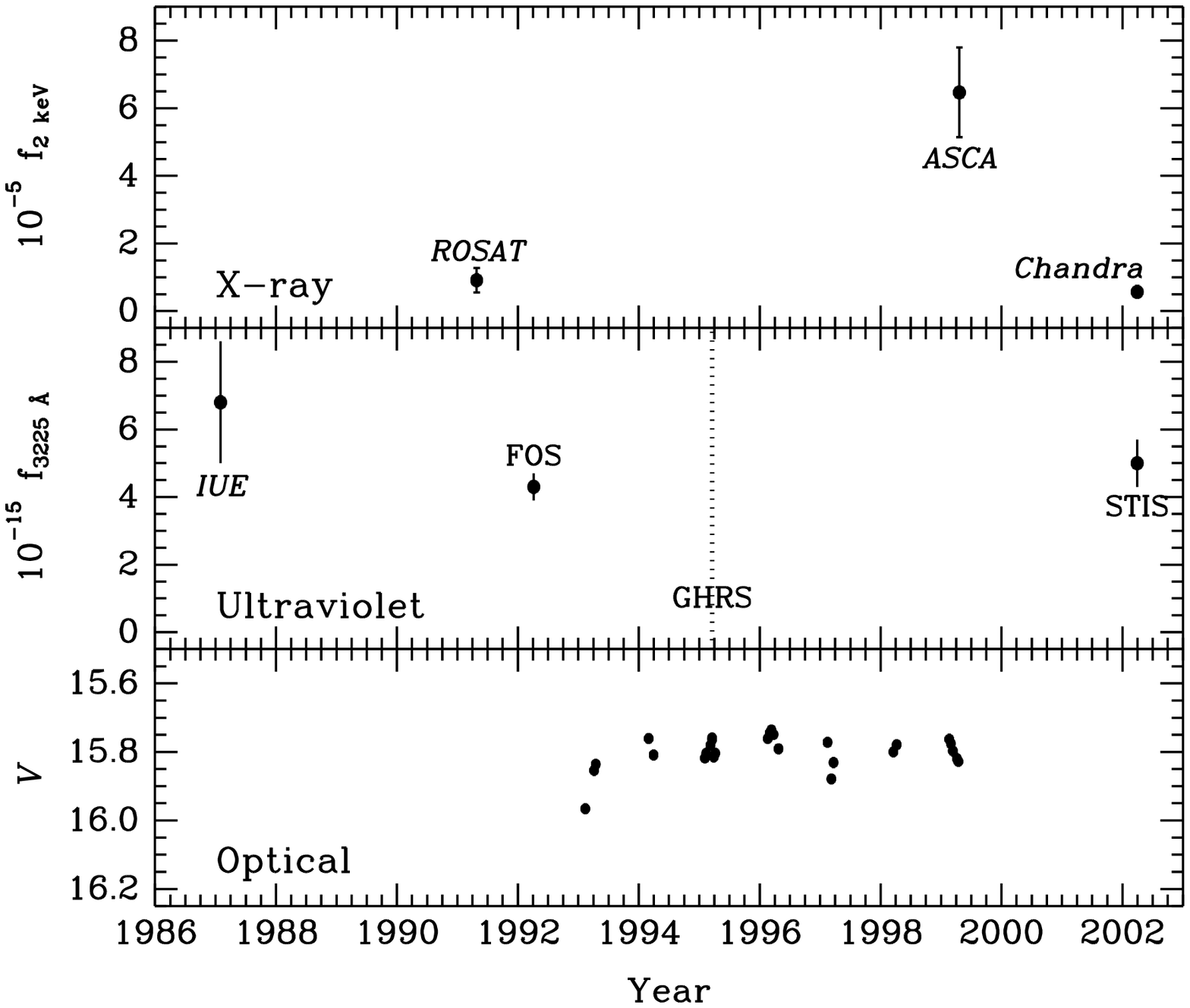}}
\caption{
Multi-wavelength light curve of \pgbal.  Data points are labelled by
observatory unless otherwise noted. {\bf Top panel:} X-ray flux
densities, in units of $10^{-5}$\fE, at observed-frame 2~keV from
model fits to the local continuum. {\bf Middle panel:} Dereddened 
UV flux densities, in units of $10^{-15}$\flam, measured at 3225~\AA\
in the observed frame (2200~\AA\ in the rest frame).  \hst\ data are indicated by labels
with the instrument name.  The GHRS observation (vertical dotted line)
did not cover this region.  {\bf Bottom panel:} Optical $V$
magnitudes from the monitoring campaign of \citet{GaRiSoTe2001}.  This
level of variability is consistent with the results from the optical
monitoring campaign of \citet{RaMcSmSt1998} who found that \pgbal\
brightened by $\sim0.18$ mag in $V$ from 1992 Sep to 1996 Jun. Note 
that the range of the ordinate represents only a factor of two in flux.
\label{fig:lc}
}
\end{figure}
\nocite{GaRiSoTe2001}
%%%%%%%%%%%%%%%%%%%%%%%%%%%%%%%%%%%%%%%%%%%%%%%%%%%%%%%%%%%%%%%%%%%%%%%%%%%%%%%%
\begin{figure}[t]
\centerline{\includegraphics[scale=0.5,angle=-90]{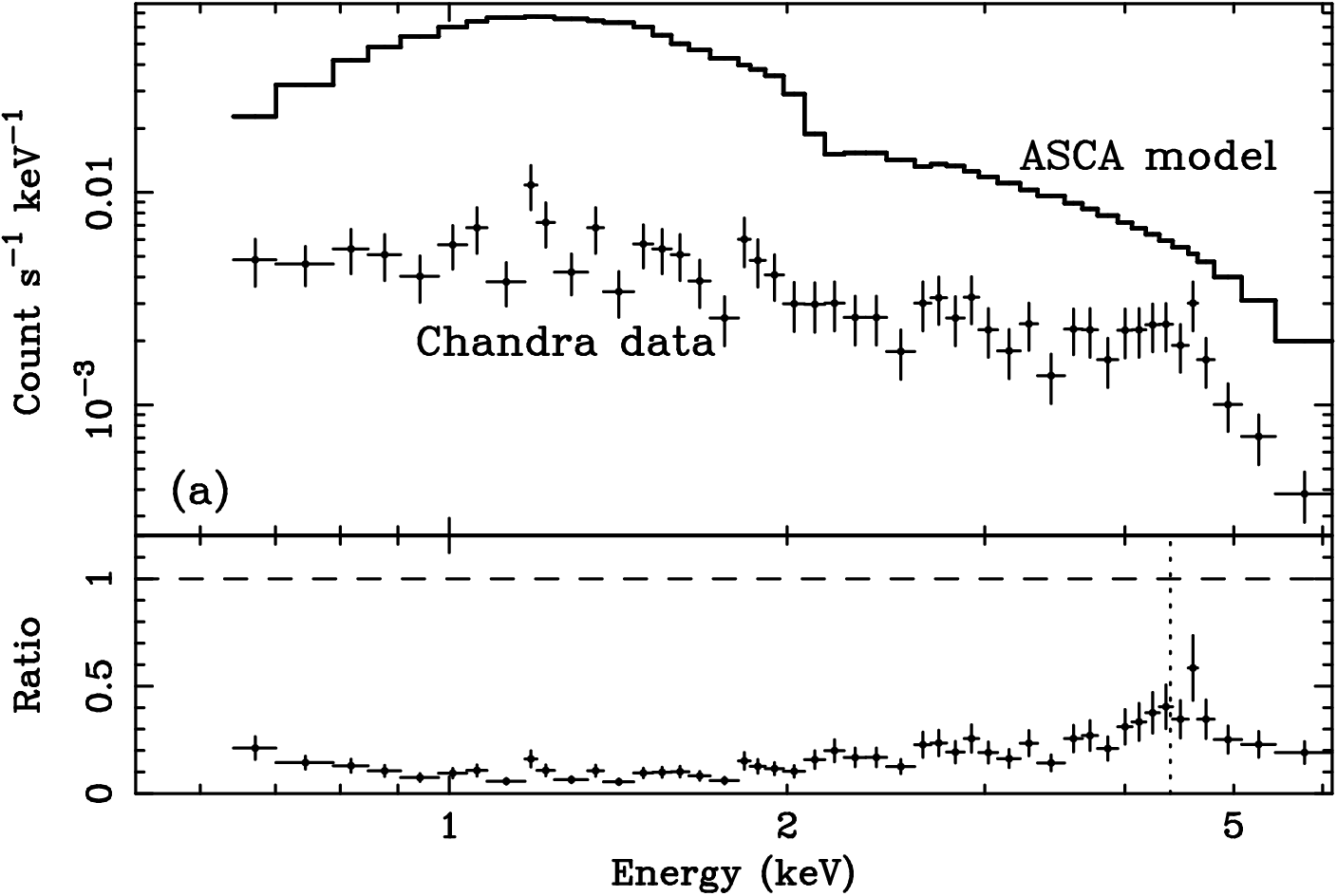}}

\centerline{\includegraphics[scale=0.5,angle=-90]{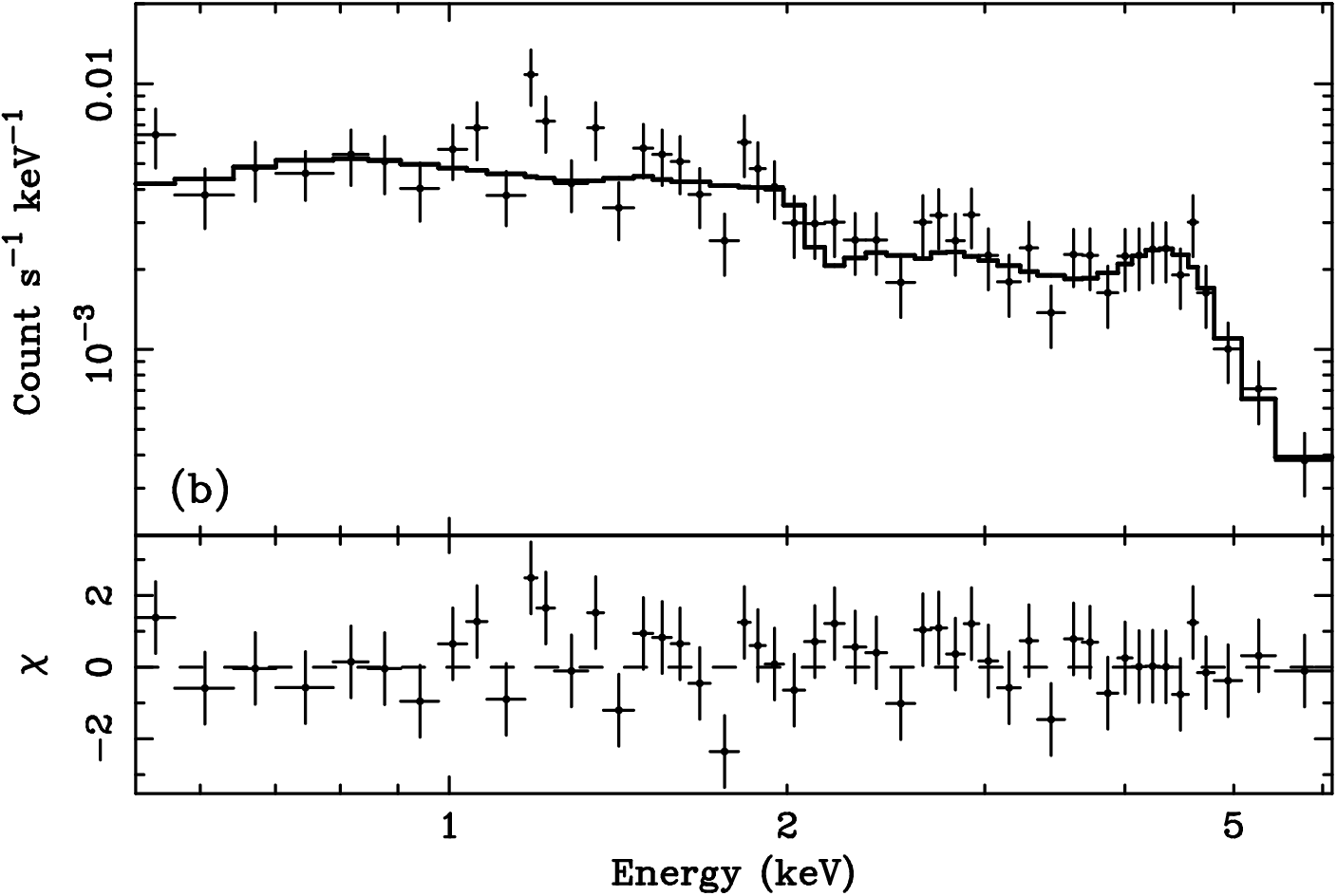}}
\caption{
Observed-frame \chandra\ ACIS-S3 spectra of \pgbal.
 {\bf (a)} The solid histogram indicates the 
best-fitting \asca\ partial-covering absorber model convolved with the 
ACIS-S3 response.  The lower panel shows the ratio of the \chandra\
data to the \asca\ model with a vertical dotted line at the energy of
the neutral \feka\ line; \pgbal\ has varied significantly in
both spectral shape and normalization since the 1999 \asca\
 observation.  The \chandra\ data have been truncated at 0.65~keV, the
 lower limit of the \asca\ data.
{\bf (b)} The solid histogram indicates the best-fitting
partial-covering absorber and broad emission-line model.  The ordinate 
for the lower panel, labelled $\chi$, shows the fit residuals in
terms of standard deviation with error bars of size unity. 
\label{fig:chandra}
}
\end{figure}
%%%%%%%%%%%%%%%%%%%%%%%%%%%%%%%%%%%%%%%%%%%%%%%%%%%%%%%%%%%%%%%%%%%%%%%%%%%%%%%%
\begin{figure}[t]
\centerline{\includegraphics[scale=0.5,angle=-90]{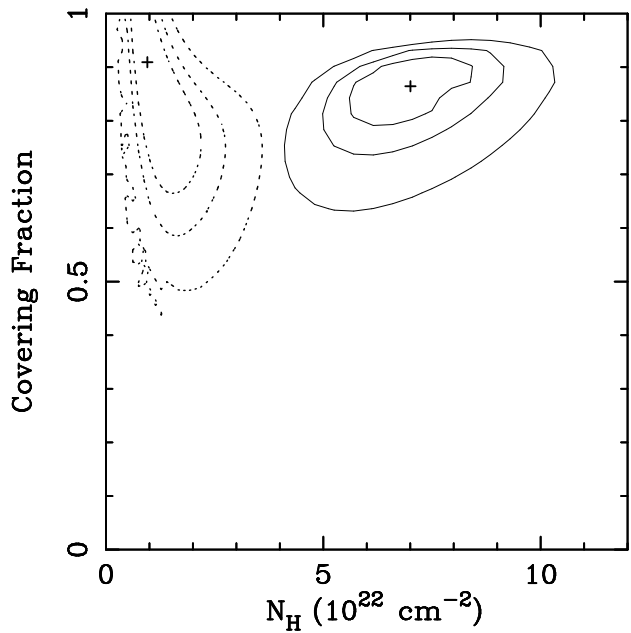}}
\caption{
  Confidence contours for the two partial-covering parameters,
  covering fraction ($f_{\rm cov}$) and intrinsic neutral column density ($\nh$), in the
  spectral analyses of \pgbal\ from the \asca\ (dotted
  contours) and \chandra\ (solid contours) observations.  The contours are
  for $68\%$, $90\%$, and $99\%$ confidence.  In both cases, the photon
  index was allowed to vary freely while calculating the contours. The
  irregularity in the $99\%$ confidence \asca\ contour at low \nh\ is likely due to the
  additional \nh\ added to the model to approximate the degradation of
  the low energy \asca\ SIS0 and SIS1 detector responses (see $\S$\ref{sec:xanal}).
  To determine the confidence contours for the \chandra\ data, the best-fitting values for
  the emission-line model were fixed.
  \label{fig:contours}
}
\end{figure}
%%%%%%%%%%%%%%%%%%%%%%%%%%%%%%%%%%%%%%%%%%%%%%%%%%%%%%%%%%%%%%%%%%%%%%%%%%%%%%%%
\begin{figure}[t]
\centerline{\includegraphics[scale=1.0,angle=0]{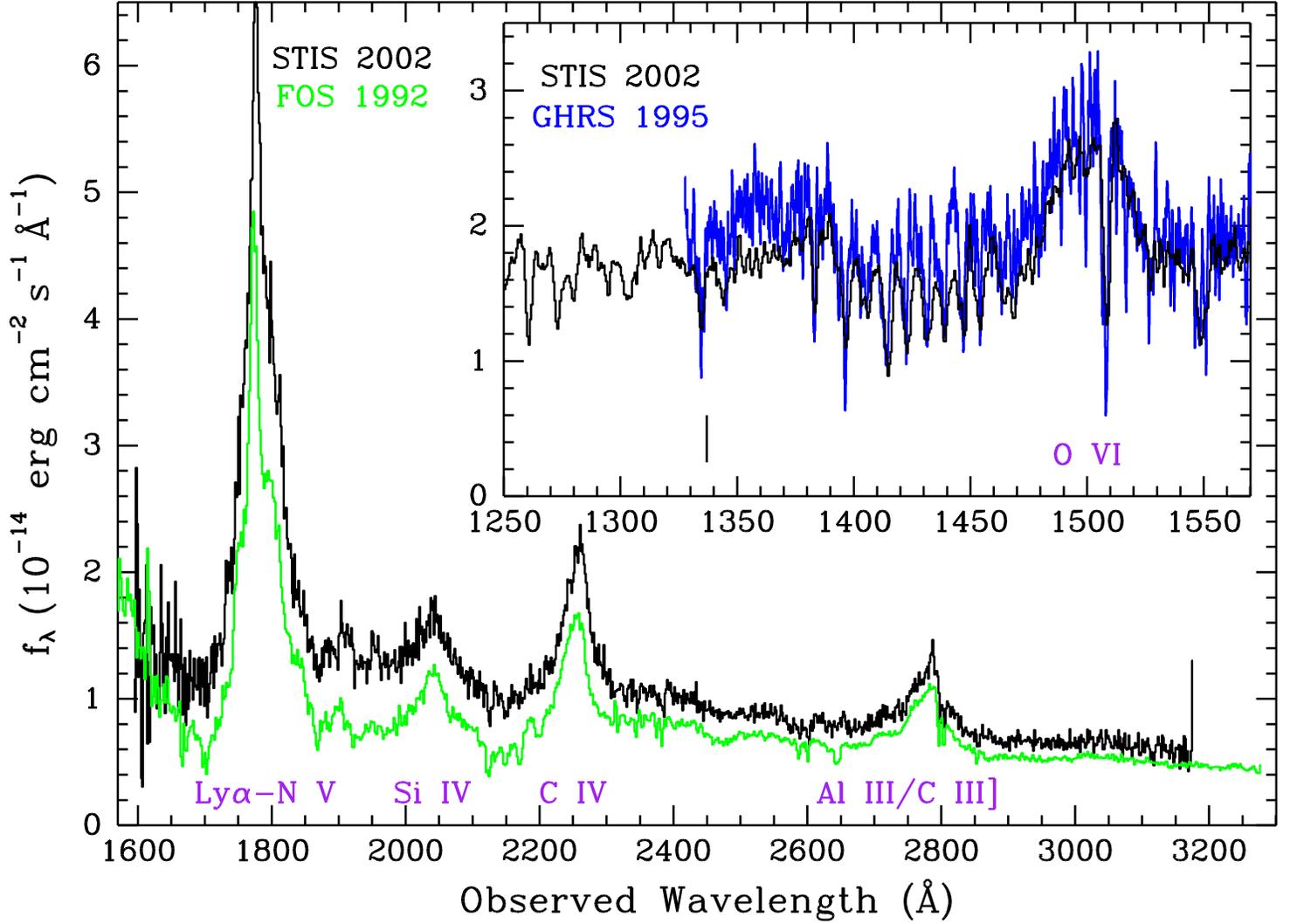}}
\caption{
Observed-frame UV \hst\ spectra of \pgbal.  Spectra in black are the
current-epoch STIS data, while those in color were previously presented in
\citet{GaEtal2001a}.  The primary broad emission lines are labelled in
purple, and the rest-frame Lyman edge wavelength is indicated with a
solid vertical line.  The spectra have been dereddened using the Galactic
\nh\,$=6.26\times10^{20}$~\cmsq\ to obtain $E(B-V)=0.125$. 
\label{fig:uv}
}
\end{figure}
%%%%%%%%%%%%%%%%%%%%%%%%%%%%%%%%%%%%%%%%%%%%%%%%%%%%%%%%%%%%%%%%%%%%%%%%%%%%%%%%
\begin{figure}[t]
\includegraphics[scale=1.0,angle=0]{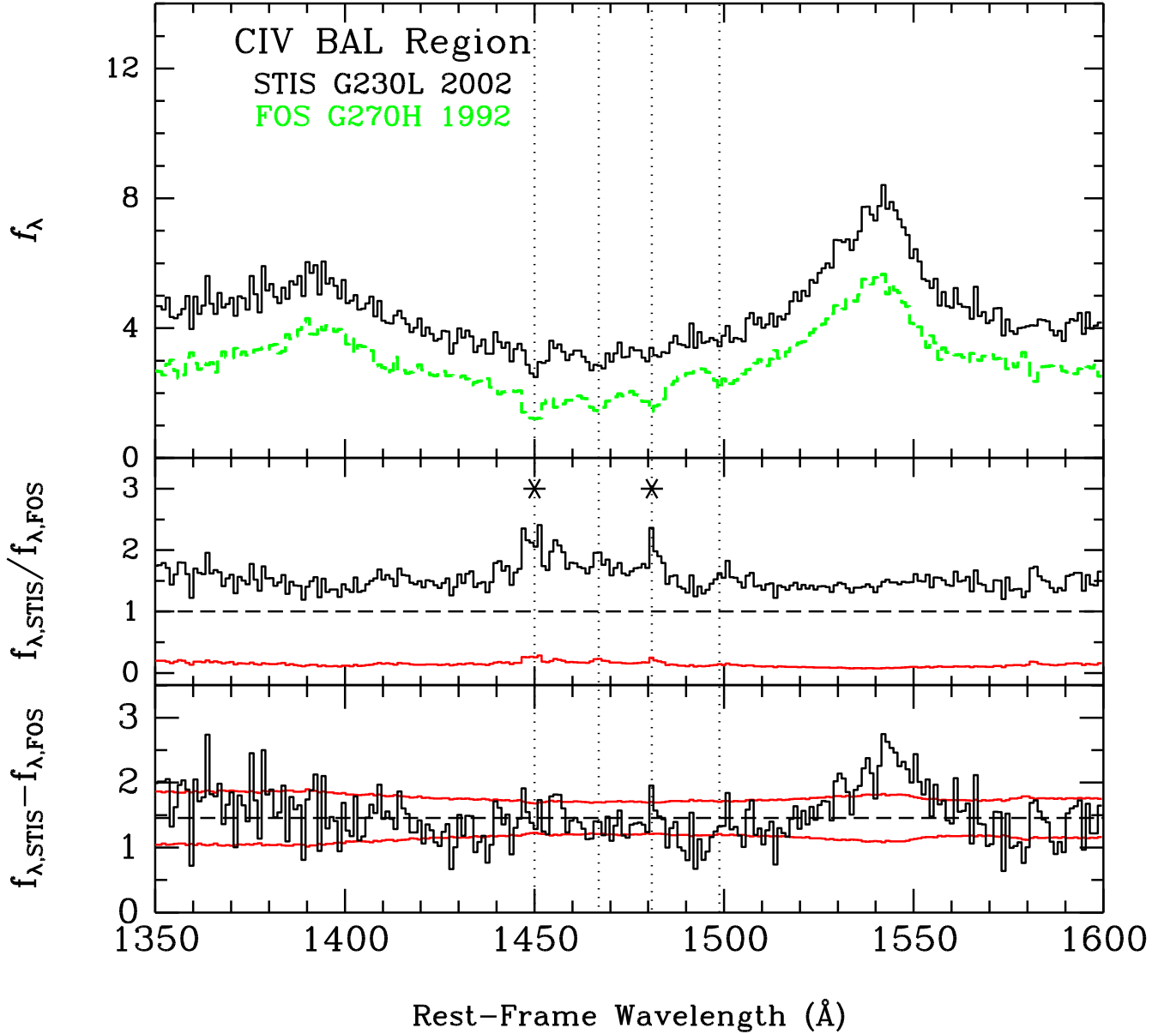}
\caption{
Enlargement of the \CIV\ BAL region.  The FOS spectrum has been rebinned to
match the lower resolution of the STIS spectrum.  Vertical dotted lines have
been drawn at the locations of absorption minima to guide the eye.
The ordinates for the top and bottom panels are in units of $10^{-15}$~\flambda.
{\bf Top panel:} The black histogram is the 2002 STIS spectrum,
while the dashed, green histogram is the 1992 FOS spectrum. {\bf Middle panel:} Division plot
with the ratio $f_{\lambda,\rm STIS}$/$f_{\lambda,\rm FOS}$ plotted at each
point.  The red histogram is the uncertainty propagated from the errors in
the individual spectra.  Asterisks mark the locations of absorption
features discussed in \S\ref{sec:UV}.  {\bf Bottom panel:} Difference plot.  The
median flux difference is indicated with a horizontal line,
while the red lines indicate the 1 sigma errors from the median.
\label{fig:c4}
}
\end{figure}
%%%%%%%%%%%%%%%%%%%%%%%%%%%%%%%%%%%%%%%%%%%%%%%%%%%%%%%%%%%%%%%%%%%%%%%%%%%%%%%%
\begin{figure}[t]
\includegraphics[scale=1.0,angle=0]{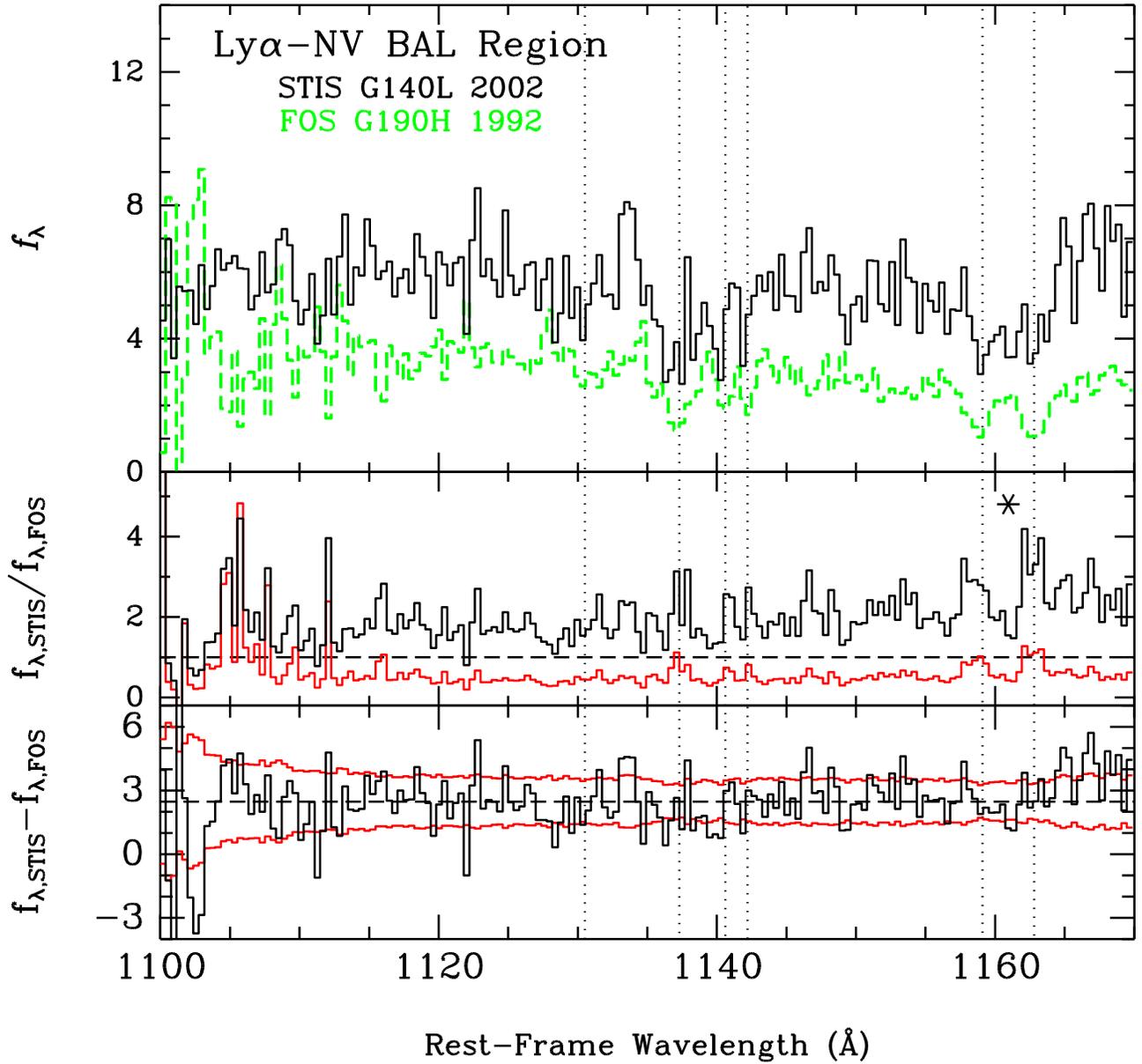}
\caption{
Enlargement of the Ly$\alpha$--\NV\ BAL region.  The FOS spectrum has been rebinned to
match the lower resolution of the STIS spectrum.  Vertical dotted lines have
been drawn at the locations of absorption minima to guide the eye.
Only the wavelength region covered by the G140L STIS observation is
included; the signal-to-noise ratio of the G270H STIS spectrum is too
low in this region.
{\bf Top panel:} The black histogram is the 2002 STIS spectrum,
and the dashed, green histogram is the 1992 FOS spectrum. {\bf Middle
  and Bottom panels:} As described in the caption to Figure~\ref{fig:c4}.
\label{fig:n5}
}
\end{figure}
%%%%%%%%%%%%%%%%%%%%%%%%%%%%%%%%%%%%%%%%%%%%%%%%%%%%%%%%%%%%%%%%%%%%%%%%%%%%%%%%
\begin{figure}[t]
\includegraphics[scale=1.0,angle=0]{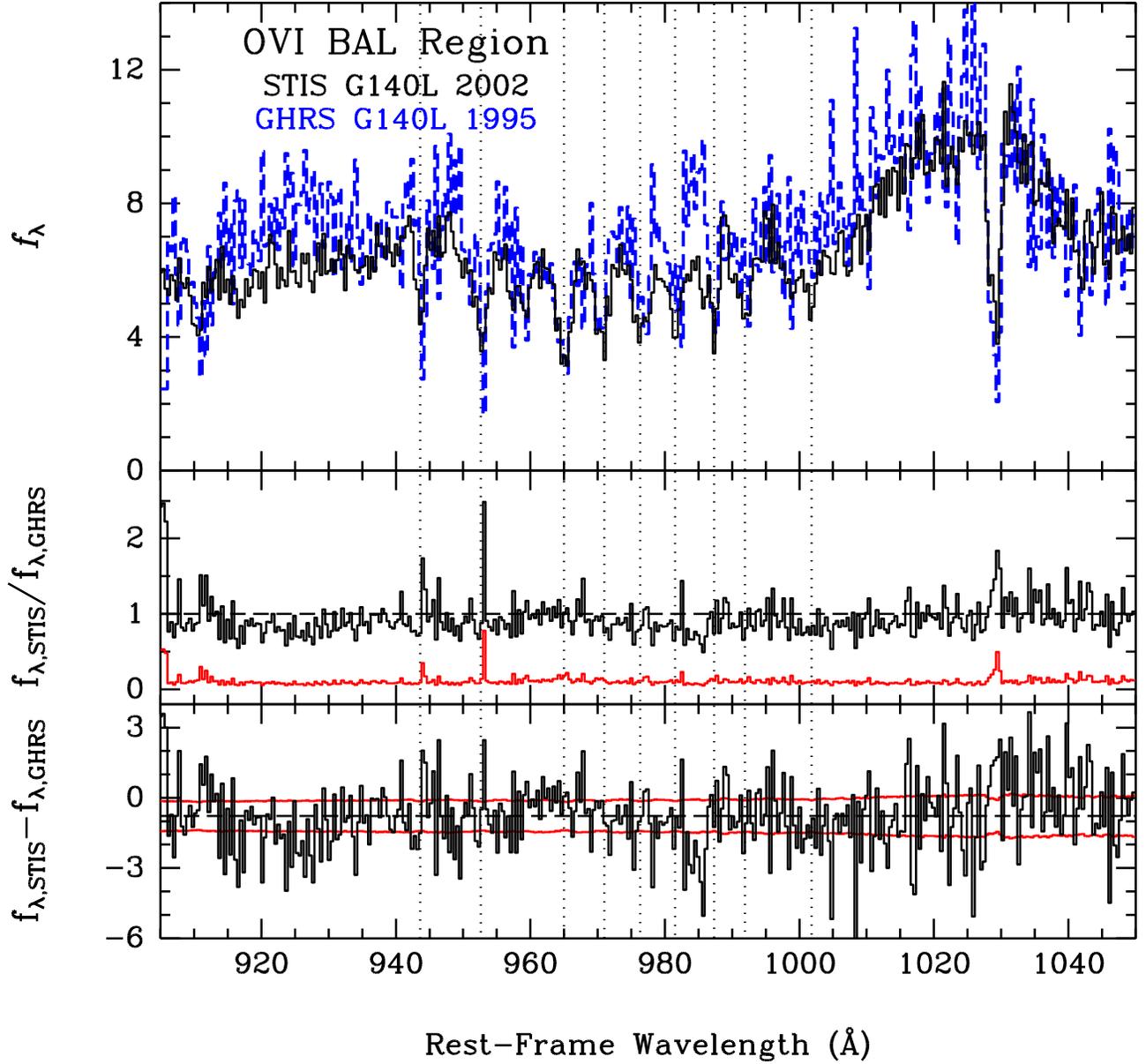}
\caption{
Enlargement of the \OVI\ BAL region.  The GHRS spectrum has been rebinned to
match the lower resolution of the STIS spectrum.  Vertical dotted lines have
been drawn at the locations of absorption minima to guide the eye.
{\bf Top panel:} The black histogram is the 2002 STIS spectrum,
and the dashed, blue histogram is the 1995 GHRS spectrum. {\bf Middle
  and Bottom panels:} As described in the caption to Figure~\ref{fig:c4}.
\label{fig:o6}
}
\end{figure}
%%%%%%%%%%%%%%%%%%%%%%%%%%%%%%%%%%%%%%%%%%%%%%%%%%%%%%%%%%%%%%%%%%%%%%%%%%%%%%%%
\begin{figure}[t]
\includegraphics[scale=0.8,angle=0]{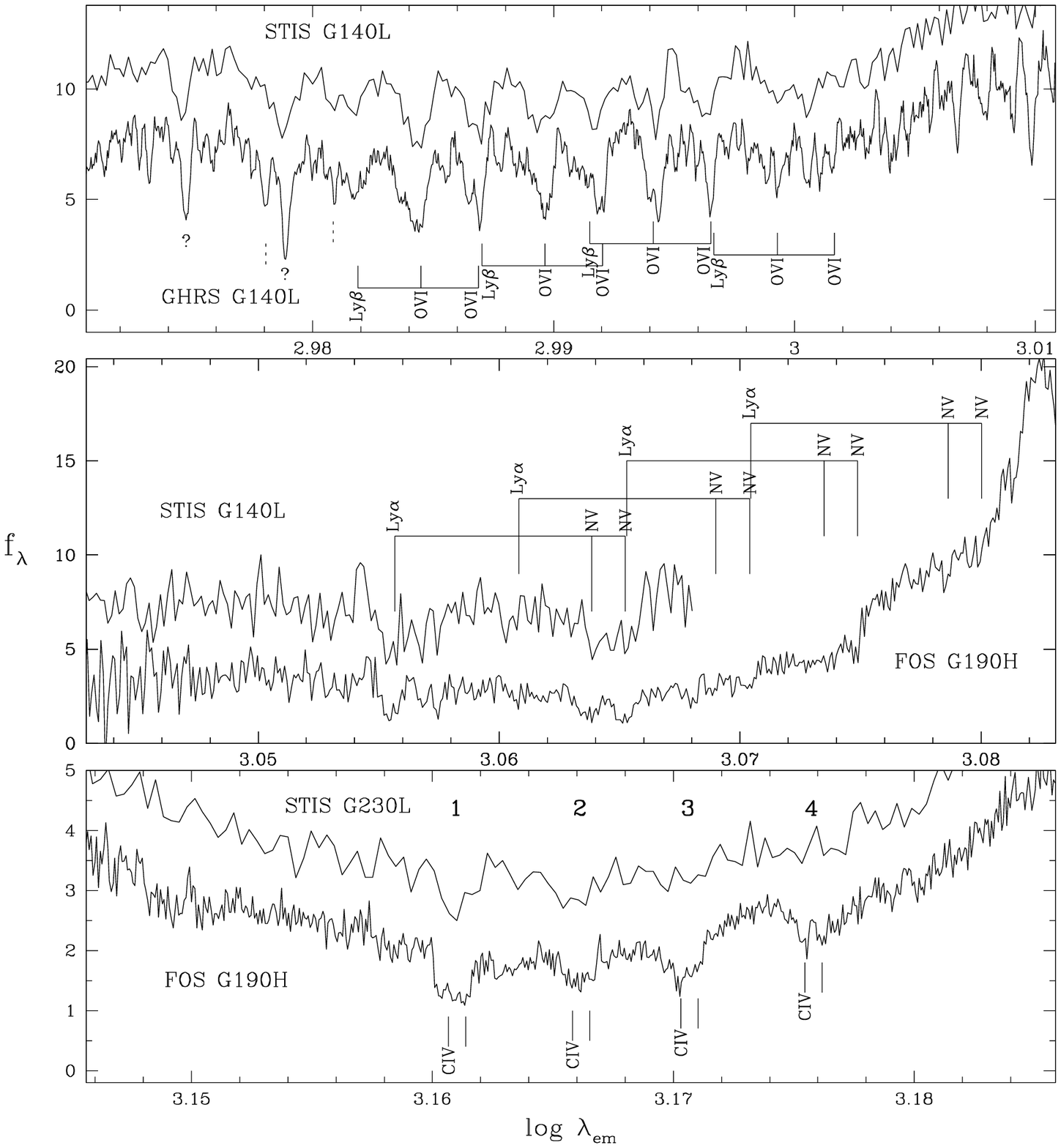}
\caption{Details of spectra from the three wavelength ranges corresponding (from
  top to bottom) to absorption of 
 Ly$\beta$ -- \OVI, Ly$\alpha$ -- \NV, and \CIV.  The ordinate in each
  panel is in units of $10^{-15}$~\flam. The data have been
plotted on the same logarithmic scale to demonstrate the redshift
coincidence of the four systems in these three wavelength regions. 
The \SiIV$\lambda\lambda1393.76,1402.77$ interstellar 
lines are indicated by dashed lines.  
There are also several unidentified lines, at rest-frame 
%observed frame from Bev.
%$1345.51$\AA$, 
%1348.98$\AA$, 
%1383.17$\AA$,
%and 1396.41 $\AA$, 
917.81~\AA,
920.18~\AA,
943.17~\AA,
and 952.53~\AA, the last two being indicated by `?'. These
unidentified lines could be
intervening Ly$\alpha$ absorbers.
\label{fig:line-locking}
}
\end{figure}
%%%%%%%%%%%%%%%%%%%%%%%%%%%%%%%%%%%%%%%%%%%%%%%%%%%%%%%%%%%%%%%%%%%%%%%%%%%%%%%%
\begin{figure}[t]
\centerline{\includegraphics[scale=0.5,angle=-90]{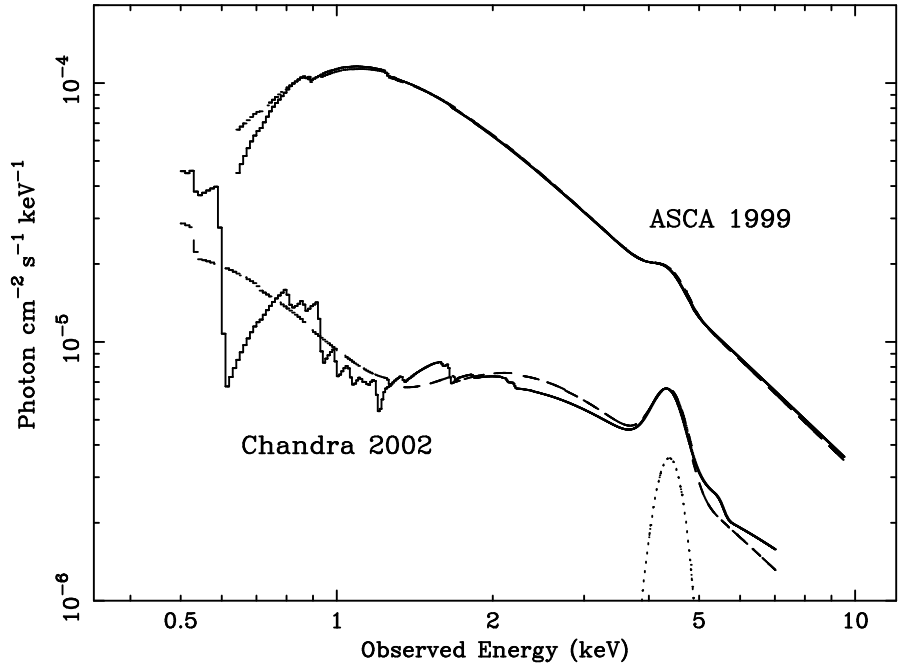}}
\caption{
Observed-frame model spectra of \pgbal\ demonstrating the spectral variability between the
1999 \asca\ and 2002 \chandra\ observations.  
The solid curves are ionized-absorber models, and the dashed curves are
partial-covering neutral absorber models.  All models also include Galactic
absorption. The emission-line model has the
parameters of the best-fitting \chandra\ model (see Table~\ref{tab1});
it cannot be positively detected in the \asca\ spectra with this normalization.
\label{fig:4models}
}
\end{figure}
\end{document}